\documentclass[twocolumn,pre,epfs,aps,showpacs]{revtex4}
\usepackage{epsfig}
\begin{document}

\title{Critical behavior of the 3D-Ising model on a poissonian random lattice}
\author{F. W. S. Lima$^1$ , U. M. S. Costa$^2$, and R. N. Costa Filho$^2$}
\affiliation{$^{1}$Departamento de F\'{\i}sica, Universidade
Federal do Piau\'{\i}, 57072-970 Teresina,Piau\'\i,
 Brazil \\$^{2}$Departamento de F\'{\i}sica,
Universidade Federal do Cear\'a, 60451-970 Fortaleza, Cear\'a,
 Brazil.}

\begin{abstract}
The single-cluster Monte Carlo algorithm and the reweighting
technique are used to simulate the 3D-ferromagnetic Ising model on
three dimensional Voronoi-Delaunay lattices. It is assumed that
the coupling factor $J$ varies with the distance $r$ between the
first neighbors as $J(r) \propto e^{-ar}$, with $a \ge 0$. The
critical exponents $\gamma/\nu$, $\beta/\nu$ , and $\nu$ are
calculated, and according to the present estimates for the
critical exponents, we argue that this random system belongs to
the same universality class of the pure three-dimensional
ferromegnetic Ising model.
\end{abstract}
\maketitle

\section{Introduction}
The Lenz-Ising model  has a vast number of applications ranging
from solid-state physics to biology. It is the oldest and most
simple model for cooperative behavior which shows spontaneous
symmetry breaking{\cite{Ising25,Temperley92}. The symmetry
breaking is very important feature of transitions that can take
place between phases with different symmetry like solid-liquid and
ferromagnetic systems. The ferromagnetic transition is an example
of continuous phase transition where the spin orientation symmetry
is broken in the ferromagnetic phase due to the formation of
magnetic domains. Inside each domain, there is a magnetic field
pointing in a fixed direction chosen spontaneously during the
phase transition.

One of the main characteristics of a continuous transition is the
existence of critical exponents in the behavior of physical
properties near the critical temperature. For example, the heat
capacity obeys a power law
\begin{equation}
C \sim |T_c - T|^{-\alpha},
\end{equation}
where $\alpha$ is the critical exponent associated with the heat
capacity. When $0 < \alpha < 1$, the heat capacity diverges at the
transition temperature. This is the behavior in the 3-dimensional
Ising model for ferromagnetic phase transition. A natural question
that follows is whether the critical exponent is modified in the
presence of impurities or a random media. In such systems the
influence of quenched disorder can be classified by the Harris
criterion\cite{Harris74,Wiscman98a,Wiscman98b,Ballesteros98} which predicts that a
perturbation is relevant if $\alpha > 0$, where $\alpha$ is the
critical exponent of the specific heat of the ideal system. For
$\alpha < 0$ quenched disorder is not relevant, while no
prediction can be made in the marginal case $\alpha = 0$.

Several studies using Monte Carlo simulations have shown that for
site-dilution and bond-dilution disorder is relevant for the 3D
ferromagnetic Ising model. However, when considering connectivity
disorder using a Voronoi-Delauney lattice in the Ising 3D model,
Janke et.al.\cite{Janke00,Janke02,Lima05}, have shown that this kind of
disorder does not modify the critical exponents. This result
contradicts the Harris criteria. The explanation is that the
disorder in the lattice was weak and more time running simulations
was necessary to obtain some meaningful shift from the critical
exponents on a regular lattice.

In this piece of work we investigate a 3D ferromagnetic Ising
model on a Voronoi lattice using the same algorithm as in Ref. 7.
But here, besides the fact that the coordination number is random,
we add an additional factor to the randomness of the Voronoi
lattice. The coupling between the neighbors spins now depends on
their relative distance $r_{ij}$ according to the expression

\begin{equation}
\label{coupling}
J_{ij}=J_{o} e^{-a r_{ij}}
\end{equation}
where $J_{o}$ is a constant and e $a\geq 0$ enters the model as a
parameter\cite{Lima00}. When $a=0$ we have the same results as in
Ref.\cite{Janke00}, i.e., the randomness in the number of
coordination does not change the critical exponents. We show here
that the addition of the new variable $J(r)$ does not change the
universality class of the Ising model.

\section{Model and Simulations}

We consider the ferromagnetic Ising model in a three dimensional
Delauney lattice where the coordination number varies locally and
the coupling factor depends on the relative distance between the
first neighbors according to Eq.\ref{coupling}. In order to build
the lattice, we first distribute $N$ sites $x_i$ randomly in a
cube with edge equal to unit. To each site $x_i$ a polyhedric
Voronoi cell $c_i$ is associated. This cell contains all the
points $x\ne x_i$ whose Euclidian distance to $d(x,x_i)$ is
inferior to the distance $d(x,x_{i+j})$. A dual lattice is then
constructed joining the $x_i$ of each cell that has a common edge.
Using such a construction, the lattice connectivity is random and
has an average value equal to $15$.

The Hamiltonian to the ferromagnetic Ising model can be described
as

\begin{equation}
\label{hamilt}
{-KH=\sum_{<ij>}J_{ij}S_{i}S_{j}}
\end{equation}
where $K=1/K_{B}T$, $T$ is the temperature, $K_{B}$ is the
Boltzmann constant, and the coupling factor $J_{ij}$ is a function
defined by Eq.\ref{coupling}. The spin $S_{i}$ can have two values
$S_i=1$ or $S_i=-1$, and the sum is evaluated in the first
neighbors only. For simplicity, the system size used is defined in
terms of a regular cubic lattice with side $L=N^{1/3}$. In our
simulations we used several system sizes;
$N=15^3,20^3,25^3,30^3,40^3,$ and $50^3$. Depending on the system
size, we have generated a different number of system copies: for
$L=40$ and $50$ we have done 25 copies; $30$ copies for $L=30$,
and $50$ copies for $L=15$, $20$, and $25$. To perform the
simulations we used the Wolf algorithm\cite{Wolff89} close to the
critical temperature $K_c$.

\begin{figure}
\includegraphics[width=8.0cm]{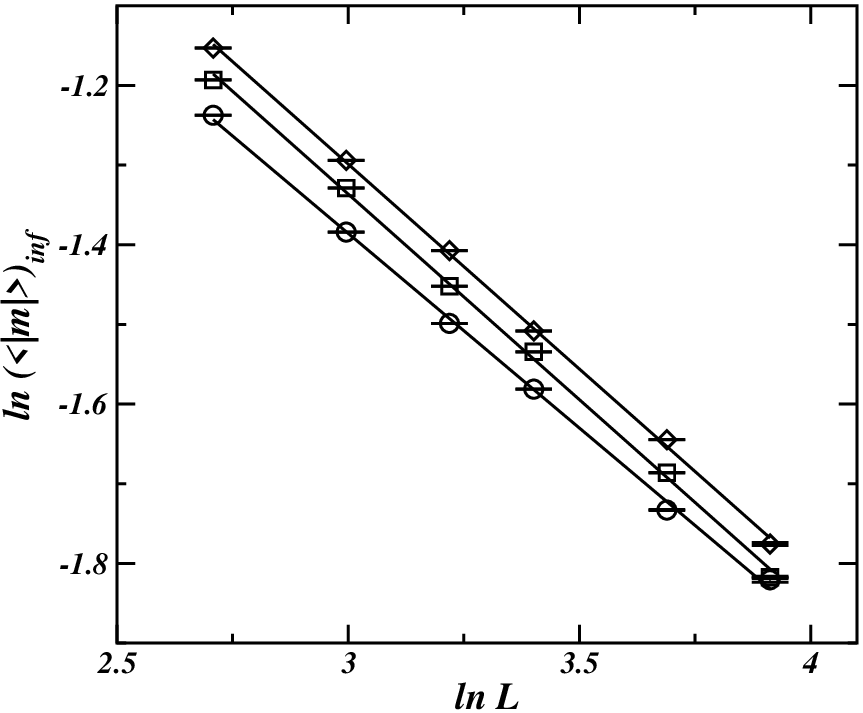}
\caption{Log-Log plot of the average magnetization $<|m|>$ versus
the system size $L$ for three different values of the coupling
parameter $a$: $a=0.0$ (circles), $a=0.5$ (squares), and $a=1.0$
(diamonds).}
\end{figure}

We start the simulations assigning to each site of the Voronoi
lattice a random spin value, i.e. the initial state is totally
random, and letting the system relax (reach equilibrium) for $120
000$ to $240000$ timesteps or clusters. After that we generate
from $2.4$x$10^6$ to $ 4.8$x$10^6$ clusters (depending on the
system size) to evaluate the averages. For each $12$ clusters the
energy per spin $e=\frac{E}{N}$ and the magnetization per spin
$m=\sum_{i}\frac{S_{i}}{N}$ were measured and registered on
temporary files. In the end, we obtained from $2.0$x$10^5$ to
$4.0$x$10^5$ values of the energy and magnetization which are is
analyzed using the histogram technique\cite{Ferrenberg98}. In
order to calculate the thermodynamic properties like specific heat
$C^{(i)}(K)=K^{2}N(<e^2>-{<e>}^2)$, for each copy classified by
the index $i$ we used the reweighting technique to calculate the
copies average indicated by the brackets in the expression below,
\begin{equation}
C(K)=[C^{(i)}(K)]\equiv\frac{1}{R}\sum_{i}^R C^{(i)}(K)
\end{equation}
The variable $R$ means the number of copies in the simulations.
The copies average is calculated over the  the $C^{(i)}$ and not
over the energy moments because the quenched averages must be
calculated at free energy level instead of the partition function
level. Finally, we determine the maximum value
$C_{max}=C(K_{max})$  for each lattice size and analyze the scale
behavior of $C_{max}$ e $K_{C_{max}}$ with the system size. Other
thermodynamic quantities like susceptibility
$\chi^{(i)}(K)=KN(<m^2>-<|m|>^2)$, magnetic cumulants
$U_{2p}^{(i)}(K)=1-(<m^{2p}>/3<|m|^{p}>^2) \;\;\;\ (p=1,2)$, and
several derivatives related to $K$, such as $d^{(i)}<|m|>/dK$, and
$d^{(i)} ln <|m|^{p}>/dK$, can be calculated in the same way. Once
the dependence on the point $K_{0}$ is well known for each run
through the reweighting, we can easily calculate the quenched
averages of the quantities.

\begin{figure}
\includegraphics[width=8.0cm]{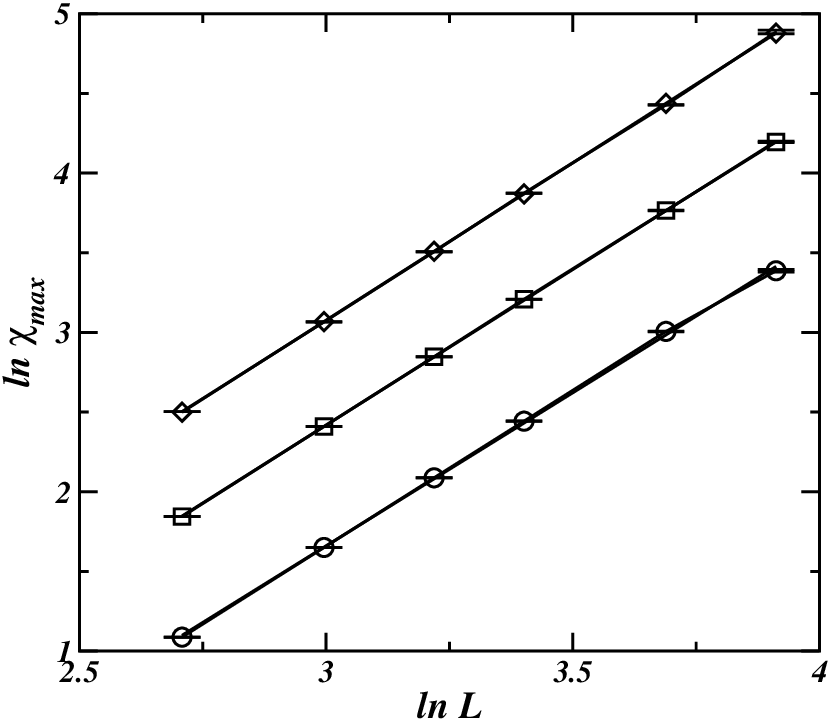}
\caption{The logarithm of the maximum susceptibility against the
system size $L$ for $a=0.0$, $a=0.5$, and $a=1.0$.}
\end{figure}

\section{Results}
To estimate the critical temperature we calculated the magnetic
cumulants $U_{2p}$ using the histogram technique in combination
with the finite size scaling analysis for three different values
of the parameter $a$ that controls the strength of the coupling
between the spins. For lattices sizes of $N\geq 8.000$ we get
$K_{c}\approx 0.0724128$ and $U^{*}_{2}\approx 0.58567$,
$K_{c}\approx 0.0724215$ and $U^{*}_{4}\approx 0.4645536$ for
$a=0.0$, $K_{c}\approx 0.139265$ and $U^{*}_{2}\approx 0.58964$,
$K_{c}\approx 0.139254$ and $U^{*}_{4}\approx 0.46976$ for
$a=0.5$, $K_{c}\approx 0.254941$ and $U^{*}_{2}\approx 0.587601$,
$K_{c}\approx 0.254953$ and $U^{*}_{4}\approx 0.46765$ for
$a=1.0$. These results show that the universal Binder cumulant
$U^{*}_{2}$ is different from $U^{*}_{4}$ for each value of the
parameter $a$ but they do not change for different values of $a$,
and the critical temperature is the same for both values of the
magnetic cumulant for each $a$. Also, the critical temperature
decreases as the parameter $a$ increases. Such behavior is very
similar to the one in dilution of pure systems.

To estimate the ratios of the exponents $\beta/\nu$ e
$\gamma/\nu$, we use the fact that the magnetization on the
inflexion point, and the susceptibility on $K^{\chi^{'}}_{max}$
scales with
\begin{equation}
         |<|m|>|_{inf}=L^{-\beta/\nu}f(tL^{1/\nu})\propto{L^{-\beta/\nu}}
\end{equation}
\begin{equation}
   \chi^{'}_{max}(L)= \chi^{'}(K^{\chi^{'}}_{max}(L),L)=AL^{\gamma/\nu},
\end{equation}
for large enough system sizes. Moreover, since the derivatives of
the magnetization with respect to $K$ depends on the factor
$L^{\gamma/\nu}$ which is an argument of the scale function
$f$\cite{Janke94},
\begin{equation}
|\frac{d<|m|>}{dK}|_{max}=L^{-\beta/\nu+1/\nu}f^{'}(tL^{1/\nu})\propto{L^{(1-\beta)\nu}},
\end{equation}
the maxima of the logarithmic derivative scales like
\begin{equation}
\label{deriv1}
|\frac{dln<|m|>}{dK}|_{max}=|\frac{d<|m|>/dK}{<|m|>}|_{max}\propto{L^{1/\nu}},
\end{equation}
and
\begin{equation}
\label{deriv2}
\left|\frac{dln<m^2>}{dK}\right|_{max}=\left|\frac{d<m^2>/dK}{<m^2>}\right|_{max}\propto{L^{1/\nu}}.
\end{equation}
The above expression determines the value of the exponent $\nu$.
Once finding that exponent, it is possible to calculate the values
$\alpha/\nu$ of the specific heat. Through the finite size scaling
analysis, it is possible to argue that the specific heat behaves
like
\begin{equation}
C_{max}(L)=C[K^{C}_{max}(l),l]=B_{0}+B_{1}lnL
\end{equation}
in the second order phase transitions. To estimate the exponents
ratio $\alpha/\nu$ we use the hiperscale relation
$\alpha/\nu=2/\nu-D$.

\begin{figure}
\includegraphics[width=8.0cm]{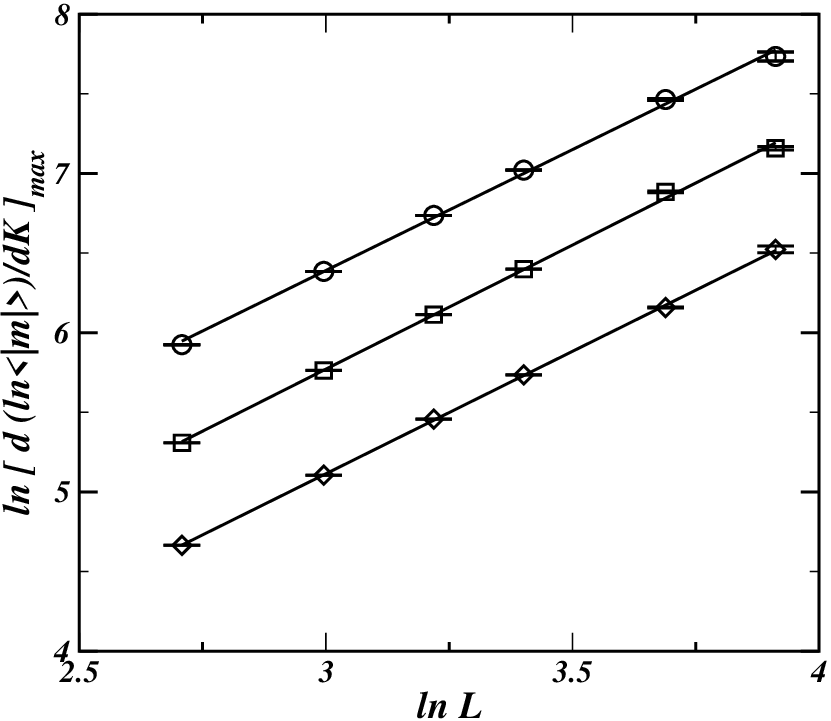}
\caption{Log-log plot of the maximum magnetization derivative
against the system size $L$ for three different values of $a$:
$a=0.0$(circles), $a=0.5$ (squares), and $a=1.0$ (diamonds).}
\end{figure}

In Figure $(1)$ we show in a logarithmic scale a graph of the
magnetization against the lattice size $L$ for three different
values of the coupling parameter $a=0.0$, $a=0.5$, and $a=1.0$.
The linear adjustment of the square minima of the simulation data
gives us $\beta/\nu=0.48920\pm0.00823$,
$\beta/\nu=0.51628\pm0.00895$, and $\beta/\nu=0.51455\pm0.00683$,
respectively. In Fig.$(2)$ we plot the logarithm of the maximum
susceptibility against the system size for $a=0.0$(circles),
$a=0.5$ (squares), and $a=1.0$ (diamonds). This curve inclination
defines the exponents ratio $\gamma/\nu=1.9223\pm0.0295$,
$\gamma/\nu=1.9537\pm0.0043$ and $\gamma/\nu=1.9758\pm0.0069$,
respectively. Figs.$(3)$ and $(4)$ show a log-log plot of the
maximum derivatives given by Eqs. \ref{deriv1} and \ref{deriv2},
respectively, against the lattice size $L$ for $a=0.0$, $a=0.5$ e
$a=1.0$. The linear fitting inclination gives
$1/\nu_{1}=1.5184\pm0.0302$, $1/\nu_{1}=1.5581\pm0.0265$ and
$1/\nu_{1}=1.5364\pm0.0095$ and $1/\nu_{2}=1.5207\pm0.0297$,
$1/\nu_{2}=1.5649\pm0.0332$ and $1/\nu_{2}=1.5326\pm0.0078$, for
the same values of $a$ as in the previous cases. In Fig. 5 we show
the maximum value of the specific heat against the lattice size
$L$ and through the hyperscale relation we get $\alpha/\nu\approx
0.0391$, $0.123$, and $0.069$, for $a=0.0$, $0.5$, and $1.0$
respectively.
\begin{figure}
\includegraphics[width=8.0cm]{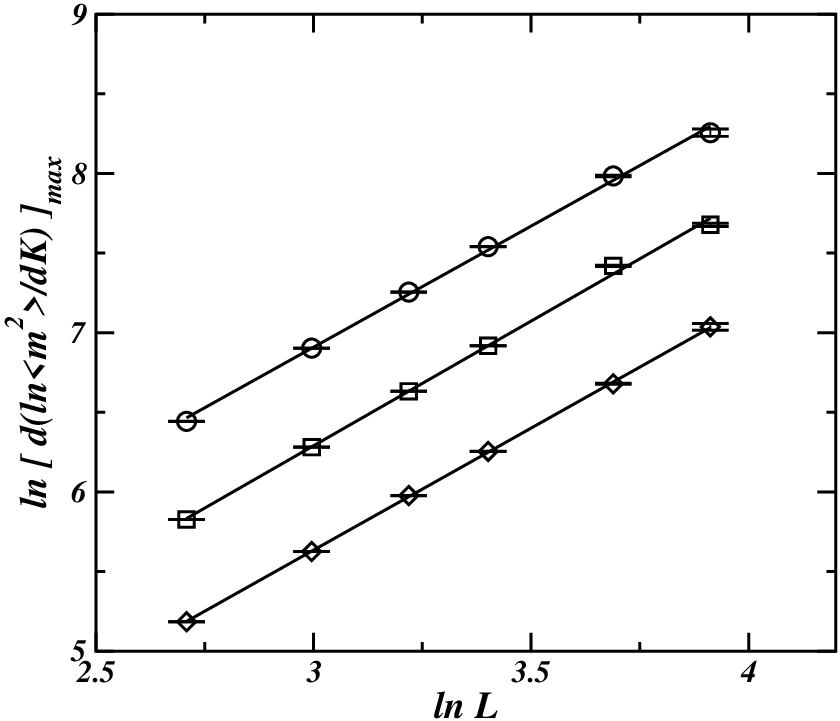}
\caption{Log-log plot of the maximum average squared magnetization
derivative (Eq.\ref{deriv2}) against the system size $L$ for three
different values of $a$: $a=0.0$(circles), $a=0.5$ (squares), and
$a=1.0$ (diamonds).}
\end{figure}
\section{Conclusion}
\begin{figure}
\includegraphics[width=8.0cm]{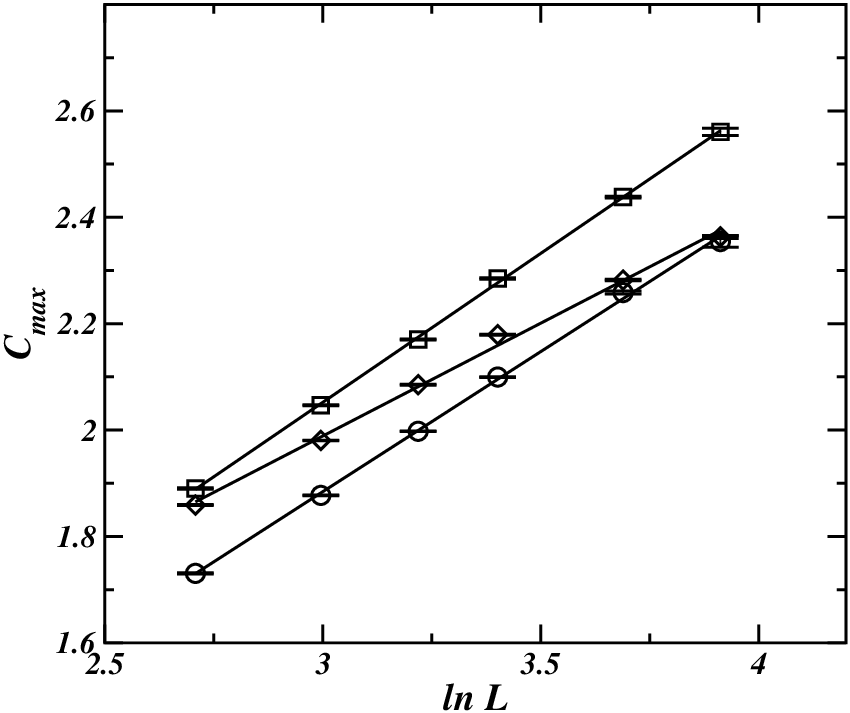}
\caption{The maximum of the specific heat against the lattice size
for different values of the parameter $a$ as in the previous
figures.}
\end{figure}

We have calculated here the critical exponents $\gamma/\nu$,
$\beta/\nu$ e $\nu$ to the ferromagnetic Ising model in a three
dimensional random Delauney lattice with a coupling factor
depending exponentially on the distances between first-neighbors
sites.

Dilution in a regular lattice was responsible for changing the
universality class of this model, while random connectivity did
not affect the exponents at all. The introduction of some kind of
dilution through a distance dependent coupling factor in addition
to the random connectivity of the lattice could increase the
strength of the quench randomness and change the exponents.
Although the critical temperature for the system decreases as the
coupling factor $a$ increases indicating a dilution effect as in
the case of a pure lattice, the results obtained here point out
that these two ingredients are not enough to drive this system
away from the universality class of the pure 3D ferromagnetic
Ising model. That can be explained by the fact that instead of
increase the randomness, the kind of dilution included here is
responsible for the weakening of the long connections. This causes
the lattice to become more regular and therefore the model is not
affected by the effective randomness introduced.

\section{Acknowledgments}

This work was supported by CNPq, FAPEPI and FAPEAL, Brazilian
agencies.

\end{document}